\documentclass{aastex63}
\newcommand{\afigtag}{-lr} 
\newcommand{\degree}{\degr}
\newcommand{\tsim}{\sim\!}
\newcommand{\masyr}{\,\mbox{mas}\,\mbox{yr}^{-1}}
\newcommand{\muasyr}{\,\mu\mbox{as}\,\mbox{yr}^{-1}}
\newcommand{\mas}{\,\mbox{mas}}
\newcommand{\muas}{\,\mu\mbox{as}}
\newcommand{\kms}{\,\mbox{km}\,\mbox{s}^{-1}}
\newcommand{\kpc}{\,\mbox{kpc}}
\newcommand{\Mpc}{\,\mbox{Mpc}}
\newcommand{\gaia}{{\it Gaia}}
\usepackage{graphicx}

\shorttitle{Gaia parallax systematics over the sky}
\shortauthors{Fardal et al.}
\begin{document}

\title{Mapping Gaia parallax systematic errors over the sky with faint Milky Way stars}

\author[0000-0003-4207-3788]{Mark A. Fardal}
\affiliation{Space Telescope Science Institute, 3700 San Martin Drive, Baltimore, MD 21218, USA}
\affiliation{Department of Astronomy, University of Massachusetts, Amherst, MA 01003-9305, USA}

\author[0000-0001-7827-7825]{Roeland van der Marel}
\affiliation{Space Telescope Science Institute, 3700 San Martin Drive, Baltimore, MD 21218, USA}

\author[0000-0003-4922-5131]{Andr\'{e}s del Pino}
\affiliation{Space Telescope Science Institute, 3700 San Martin Drive, Baltimore, MD 21218, USA}

\author[0000-0001-8368-0221]{Sangmo Tony Sohn}
\affiliation{Space Telescope Science Institute, 3700 San Martin Drive, Baltimore, MD 21218, USA}

\correspondingauthor{Mark Fardal}
\email{fardal@stsci.edu}

\begin{abstract}
Parallaxes measured by the \gaia\ mission have huge significance for astronomy, 
  but parallaxes in \gaia\ DR2 are known to have systematic errors
  that depend on the source position and other quantities.
We use the abundant information in faint Milky Way stars,
  along with the GOG simulation of the \gaia\ catalog,
  to probe the spatial dependence of \gaia\ DR2 parallax systematic errors
  in an empirical way.
The parallax signal, concentrated in thick disk turnoff stars
  with magnitude $G \sim 17$, is sufficient to
  construct maps of the parallax systematic error over the majority of the sky.
These maps show
  a locally regular ``waffle pattern'' on $\tsim 1 \degree$ scales following \gaia\ scan directions,
  stronger linear ``scar'' features,
  and coherent variations on larger scales.
The parallax bias maps also retain traces of astrophysical effects such as dust clouds.
The waffle pattern, known from earlier maps of the Magellanic Clouds,
  extends over the entire sky; its local rms amplitude averages $15 \muas$
  and varies by about a factor of two.
The strength of this pattern increases by a factor $\tsim 6$ 
   from magnitude $G = 13$ to $G = 20$.
Correlations with parallaxes of quasars and of stars with independent distance estimates
   support our bias estimates.
Using similar methods, we map systematic errors in the proper motion
 and examine the relationship with the parallax systematics.
We provide a code package to access and query our bias maps.
Similar tests on the general stellar population should be useful 
in quantifying systematic errors in future \gaia\ releases.
\end{abstract}
\section{Introduction}
\label{sec.intro}
With its Data Release 2 (DR2), the \gaia\ satellite has transformed
our view of the Milky Way stellar population.  In large part this is due
  to its astrometric data, which provides proper motions (PM) and
  parallax for a huge number of stars over the entire sky, and
  has yielded a rich bounty of scientific results too numerous
  to list here.  
Precisely because of the extraordinary size and impact of this dataset, it is
important to understand its systematic errors,
to help sort measurements that can be accurately made with \gaia\
from those that cannot.  
Much work has already been done on biases in the parallax measurements, beginning
with the DR2 release papers \citep{lindegren18,arenou18,mignard18,helmi18},
aided by comparison to various types of sources.
Quasars, which in reality have negligible parallax, were found to have
a definite zero-point parallax bias.
Furthermore, the parallax and proper motion biases
show significant variation over the sky, as demonstrated both
with quasars and with the densely concentrated stars of the Large and Small Magellanic Clouds (LMC and SMC).
Subsequent studies addressed the parallax systematics in other types of sources, 
  including Cepheids, eclipsing binaries, and asteroseismic targets 
  \citep[e.g.][]{riess18,stassun18,zinn19,khan19}.
Much of this later work focuses on establishing an overall parallax bias level
for the particular sample under study.
However, the significant parallax bias variation over the sky
  can complicate comparisons between samples
  (especially those based on small patches of the sky, such as the Kepler field targets)
  and attempts to remove bias.  Several papers have generated sky maps of 
  the bias \citep[e.g.\ \gaia\ papers,][]{khan19,chan20},
  albeit at relatively low resolution.  

In DR2, systematic errors in the parallax and proper motion
  are negligible compared to statistical errors for individual sources.
For coherent groups of stars, however, systematics can be the dominant error.
For example, \citet{helmi18} measured proper motions for a sample of globular
  clusters but needed to assume literature values for 
  the distances, since the systematic errors from \gaia\ exceeded both the
  tiny \gaia\ statistical errors and reasonable errors on the literature distances.
For Local Group galaxies at a distance of $\tsim 1 \Mpc$,
a typical proper motion systematic of $\tsim 70 \muasyr$ \citep{lindegren18}
corresponds to a velocity error of $> 300 \kms$, larger than the expected
physical signal.  
Upcoming \gaia\ data releases starting with Early Data Release 3 (EDR3) should improve on the systematics.  
At this point, though, it is unclear whether systematics will diminish faster,
  at the same rate, or slower than the statistical errors in future releases.
It would be very useful to have a way to test the 
  parallax and proper motion biases in EDR3 and future releases.

The principal goal of this paper is to use the large numbers of Milky Way 
stars in \gaia\ DR2 to probe the systematic errors in the parallax.
A secondary goal is to improve our understanding of
systematic errors in the proper motion, using similar methods as for the parallax.
One might reasonably ask whether our assessment of systematic
biases could be translated into a method for correcting them.
For now, we must urge caution about doing so, since we find that the
estimates of parallax or proper motion bias at any particular point
have significant errors of their own. Thus this technique is best
used to find the limits of applicability of the \gaia\ data,
in DR2 and later releases.

The plan of the paper is as follows:
  in Section~\ref{sec.datamethods} we first describe the datasets used, then our
  methods for constructing maps probing the parallax bias using \gaia\ catalog stars.
We describe our main results
in Section~\ref{sec.results} with a series of all-sky maps and tests of 
  correlations with other samples.
We compare analogous proper motion systematics in Section~\ref{sec.pm},
  and present our overall conclusions in Section~\ref{sec.conclusions}.

\section{Data and methods}
\label{sec.datamethods}
\subsection{Data sources}
\label{sec.data}
In this paper we use astrometric and photometric parameters from \gaia\ DR2.
These are obtained
over the entire sky for $G \lesssim 20$, with a gradual cutoff at the faint end.
We make no use here of the radial velocities or stellar parameters
\gaia\ provides for a subset of sources.  
Information on the \gaia\ mission generally and DR2 specifically
can be found in \citet{gaiaMission}, \citet{brown18}, and associated papers.

We use a sample of sources downloaded from the \gaia\
European Space Astronomy Centre (ESAC) server using the
table access protocol (TAP).
We use a sequence of queries targeting specific HEALpix cells
to cover the required region of the sky.
We limit our sample to fainter stars with $G > 13$,
since at brighter magnitudes the
technique used for astrometry differs \citep{lindegren18}
and the systematics may therefore be quite different.
We retain only sources with measured parallaxes and $BP - RP$ colors.
To avoid extreme outliers that could contaminate the statistical
variations we seek, which are at the level of tens of $\muas$,
we exclude sources with parallax greater than $5 \mas$.
We also cut the sample using \gaia\ astrometric quality indicators,
keeping only sources with renormalized unit weight error (RUWE)
less than 1.5 and astrometric excess noise (AEN) less than 4.0.
While the quality indicators do show systematic variations on the sky,
the parallax bias indicators we construct later do not appear
to be affected by this cut much if at all, wherever we have compared.

We found it helpful to analyze the sample in small portions of the sky.
Our main data sample is made up of overlapping circular fields of radius $6 \degree$,
where the centers are placed using the level-3 HEALpix scheme in Galactic coordinates.
We exclude the densest regions near the Galactic bulge and midplane.
This leaves 590 fields, half in each Galactic hemisphere.  

We use the Gaia Object Generator (GOG) simulation \citep{luri14} as a control sample.
This is an updated version of the original Besan\c{c}on model \citep{robin03,robin12} 
with realistic measurement errors added.
Since the TAP server was offline during the period when we were obtaining the data,
we instead downloaded the files containing the entire simulation,
and reorganized the contents into smaller files for quicker access.

We also use three auxiliary samples for testing the parallax bias.  
The first is the \citet{hogg19} sample of red giant branch (RGB) stars,
for which distance was estimated from spectroscopy.
These sources are distributed unevenly across the Northern sky,
and many of them are in the bulge and disk regions we exclude from consideration.
The second is the red clump (RC) star catalog
in the Kepler field from \citet{hall19},
where the source selection is made using asteroseismic measurements.
The distance to these sources can be estimated with both asteroseimology
and photometry.
The third sample is the quasars from \citet{secrest15},
which we obtained from the table {\tt aux\_allwise\_agn\_gdr2\_cross\_id}
in the \gaia\ DR2 database.
These mostly faint and blue sources should have negligible true parallax.  
Conveniently, the sky distribution of this sample continues just far enough into the
Galactic plane to cover our chosen set of fields.
We exclude these quasars from our Galactic star samples.
This has negligible effects on our main results since they represent
a small fraction of the total sample,
but is important when cross-correlating with the quasars themselves.
\subsection{Local parallax bias maps}
\label{sec.methods}
Our analysis is primarily based on the $6 \degree$-radius fields defined above,
processed individually at first.
These fields are large enough to show the repeating degree-scale
``waffle pattern'' previously shown in,
e.g., the LMC \citep{lindegren18,helmi18},
  while small enough to make the computational burden and
  effects of population gradients manageable.
We assign stars to spatial pixels 
using tangent-plane coordinates $\xi$, $\eta$ around the field center
which are aligned with equatorial coordinates.
  
\begin{figure}[t!]
\centerline{\includegraphics[width=152mm]{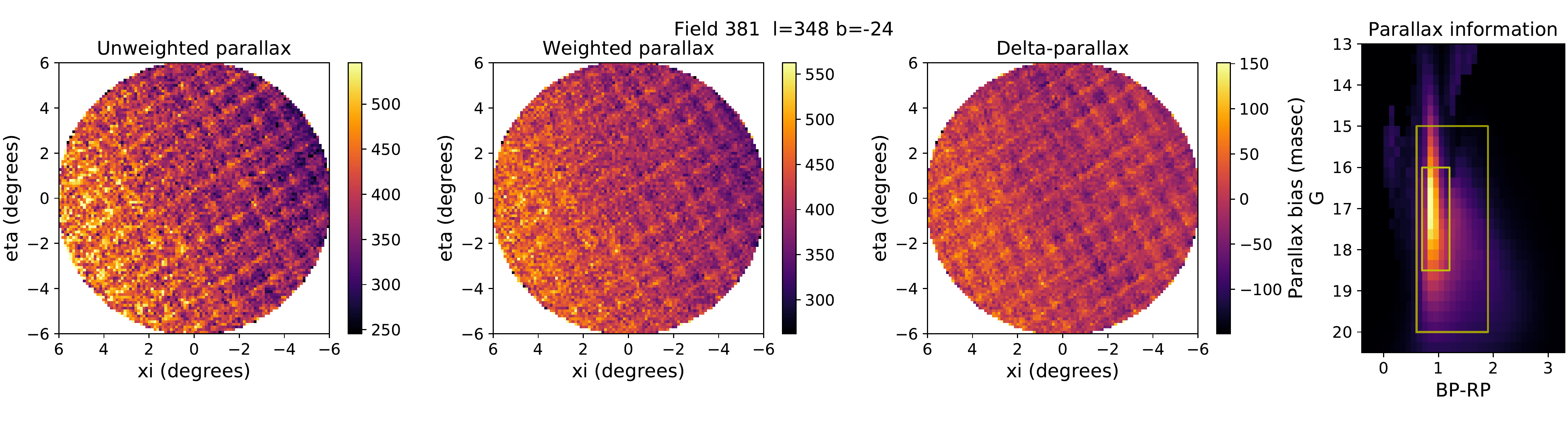}}
\centerline{\includegraphics[width=152mm]{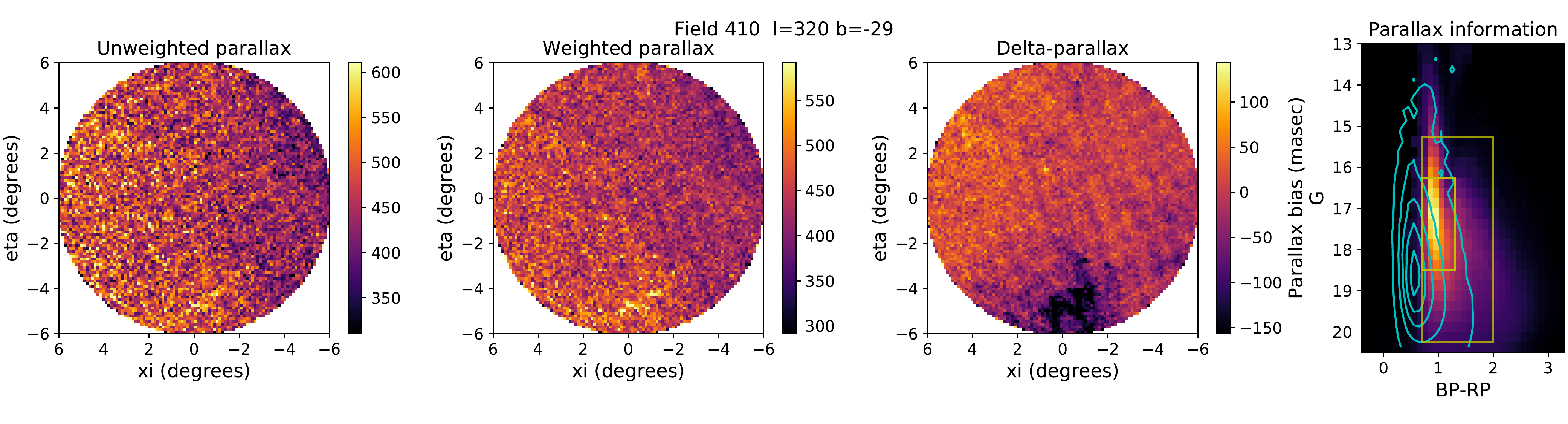}}
\centerline{\includegraphics[width=152mm]{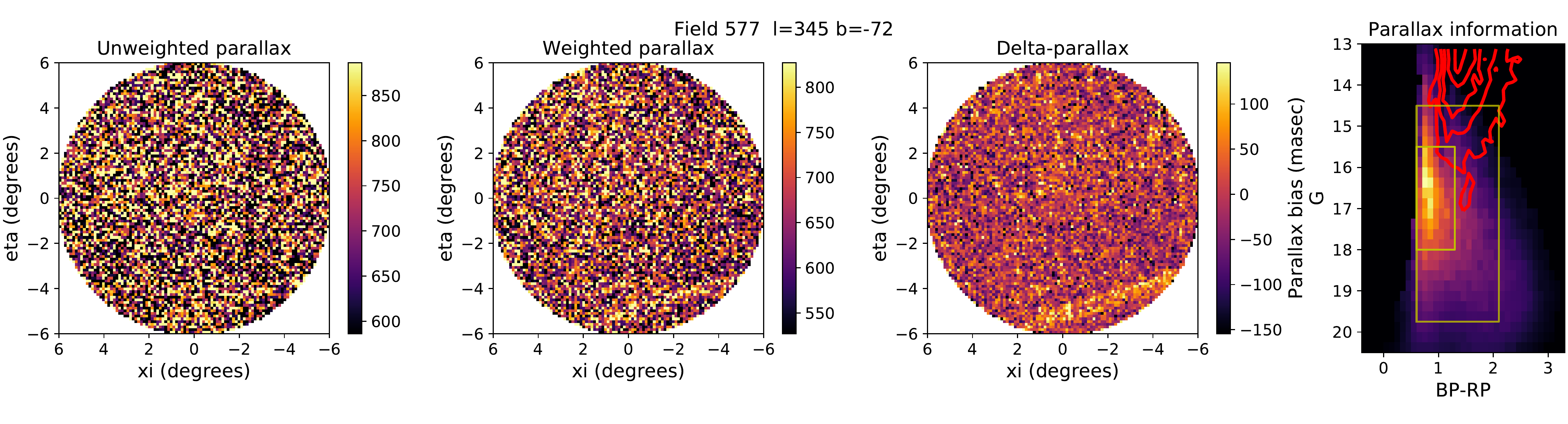}}
\caption{
\label{fig.examplemaps}
Example bias maps in three different fields.
The algorithms used are the median parallax, weighted average parallax,
and weighted average delta-parallax (difference from expectation
based on color and magnitude).
The colorbars show the scale in units of $\muas$.
These algorithms are explained in more detail in the text.
Map axes $\xi$, $\eta$ are a tangent-plane projection aligned with
equatorial coordinates, with north at the top.
Stellar density decreases from the first to the third row.
The first row is an especially clear example of the
``waffle pattern'' systematic.
The second row shows the influence of a dust cloud
in the second and third columns in the light and dark regions at the bottom
of the field.
The third row shows a strong ``scar'' feature at the bottom of the field.
The final column shows the
information (sum over source weight) in the color-magnitude diagram.
Brighter colors indicate more information, and 
the boxes contain 90\% and 50\% of the total information.
The contours in the second row show the information in the full quasar
sample, and those in the third row show the information in the
sample of RGB with $|b| > 10 \degree$.
}
\end{figure}

We have tested several ways to probe the spatial variation of the parallax bias 
in the DR2 sample.  
The simplest way is to compute the median parallax of all sources in each
spatial pixel. 
This is effective in the LMC or SMC, due to the high source density and
a true parallax that is both small and uniform.
We show maps constructed this way in three different fields in the first column
of Figure~\ref{fig.examplemaps}.  In the first row the stellar density is
relatively high and the mean parallax is fairly smooth, with the exception of a
clear waffle pattern.  In the second and third rows, 
the stellar density decreases and the statistical noise increases as a result.

One factor that increases the noise in these maps,
relative to maps of the LMC,
is that stars in any given spatial pixel have a broad range of true
parallax values, with distributions that relate to their color and brightness.
We therefore found it useful to weight sources according to their position in
the color-magnitude diagram (CMD).
The information about parallax bias contributed by any particular source is controlled
by the spread in observed parallax,
  which combines the effect of observational error (rising towards
  faint magnitudes) and intrinsic parallax variation
  (generally rising towards bright magnitudes).
The information contributed by any bin in the CMD is a product of the per-source
  information and the stellar density within that bin.

We constructed weighted-average parallax maps with the following algorithm.  
We bin the stars in the CMD, using 20 bins in color over $0.5 < BP - RP < 3.5$,
and 35 bins in magnitude over $13.0 < G < 20.0$.
In each CMD pixel, we measure the median parallax and the parallax spread,
defined as the interquartile range (25th to 75th percentile) rescaled
so it matches a $1 \sigma$ spread for a Gaussian distribution.
These quantile-based measures are aimed at reducing the influence of
non-Gaussian tails of the parallax distribution.
We exclude sources with parallax differing from the median value
by more than 4 times the parallax spread.
We also limit consideration of CMD pixels to those with more than 40 sources,
to avoid poorly defined distributions.
We then weight individual sources by the inverse square of the spread.
Finally, we compute the weighted mean parallax from the retained sources in
each spatial pixel.
This averaging method does
make the physical meaning less clear,
and also somewhat variable since the weighting scheme varies over the sky.  
However, it reduces the random noise
compared to a plain average of the parallax.
Maps constructed by this method are shown in the second column of
Figure~\ref{fig.examplemaps}.  Especially in the second and third
rows, they display less noise on a pixel scale than
the unweighted maps in the first column.  As a result more features
can be distinguished, including a bright spot at the bottom of the second
field and a bright stripe at the bottom of the third.  

We also construct ``delta-parallax'' maps, which represent
the {\em difference} of the stellar parallax from its expected value.
In this case, the quantity we average is the source parallax minus
the median parallax for sources in its CMD pixel.
Otherwise, this is calculated like the mean parallax map 
using the same weighting scheme.
These maps suppress noise induced by the random sampling of different 
parts of the CMD.
The delta-parallax maps are by nature close to zero when averaged over the field,
since we are calibrating the \gaia\ sources within the field against themselves.

Maps constructed with this method are shown in the third column of
Figure~\ref{fig.examplemaps}.  The pixel noise is now further suppressed
compared to the two previous methods.
Many of the features are similar between the second and third columns,
including the general waffle pattern and the bright linear
feature in the third row.
However, in the middle row, we can see a feature due to dust,
which shows up as a bright (high) offset
in the parallax map and a dark (low) offset in the delta-parallax map.
This pattern is typical of dense dust clouds in our maps, and is
an artifact of our method rather than \gaia.
In the parallax map, dust clouds block the more
distant stars and leave the nearer ones with high parallax,
biasing the average upwards.
In the delta-parallax map, which compares the parallax to the expectation
based on CMD location, that selection bias is reduced.
Instead, reddening due to the dust moves stars within the CMD,
causing them to be compared
to nearby stars and giving a bias in the opposite direction.
Primarily because of these effects of dust on our maps, we use both
parallax and delta-parallax maps as probes of the \gaia\ systematic error.

An implicit assumption behind our inverse-variance weighting
scheme is that the spatial dependence of the parallax bias
is separable from any dependence on color
and magnitude, as well as other possible factors like crowding.
While this assumption is a useful starting point,
we will later see it is {\it not} quite true.
Thus, the following maps and associated colorbar scales represent the 
local average of the parallax bias in our information-weighted scheme, but not
necessarily the parallax bias of any given source.  

We construct analogous parallax and delta-parallax maps
using the GOG simulation.
While overall GOG is an impressive prediction of the \gaia\ source properties,
there are clear differences.
The CMD distribution of the GOG sources displays discrete stripes at particular colors.
The mean color of the turnoff is slighly offset to the blue compared to \gaia\ data
in most fields we have examined.
Furthermore, there are offsets in the parallax and delta-parallax maps
from the observed maps, which vary over the sky.
Some of this is due to \gaia\ systematic errors previously reported
in the literature.  But we also find the 
mean simulated parallax in a given CMD pixel near the main-sequence turnoff
is typically 10--20\% higher than in the observed sources.
These differences can be much larger than expected \gaia\ systematic errors,
and suggest a systematic offset in source brightness of roughly 0.4 magnitudes,
perhaps from systematic differences in the age and composition 
of the simulated and observed stars.
We have not made any attempt to adjust the GOG source parallax prior
to constructing the maps.
The statistical errors in GOG are intended to match the predicted
\gaia\ end-of-mission properties.
However, at the faint end the errors are actually quite similar to DR2, while at the
  bright end the spread is dominated by intrinsic variation in parallax rather than
  statistical errors.
We therefore used the GOG ``observed'' parallax values as given, 
  rather than trying to make the statistical errors resemble DR2 more closely.

GOG includes a model of Galactic dust, so parallax maps constructed
from it reproduce most of the dust features in \gaia.  However, we
found that the GOG dust features were typically somewhat more extended and
diffuse than the observed features.  We obtained closer matches
to the \gaia\ maps by replacing the extinction values within GOG with
values calculated from the maps of
\citet{schlegel98}\footnote{We used the Python implementation of the Schlegel et al.\ maps in {\tt github.com/kbarbary/sfdmap}.}, and adjusting the source
magnitudes and colors accordingly.  We have not adjusted the other
source parameters to account for this change, as the error properties
change slowly with magnitude.  

In constructing simulated maps, we use the \gaia\
sample in the same field as a reference to determine the 
source weighting and median parallax in the CMD,
using GOG only for the parallax values that are averaged.
The difference between \gaia\ and GOG delta-parallax maps thus 
indicates the observed parallax bias of \gaia\ sources, combined with any
systematic error in the GOG model itself.
Later we will examine the extent to which we can decompose those two factors.
These four types of maps in the $6\degree$-radius fields --- observed parallax,
observed delta-parallax, simulated parallax, and simulated delta-parallax --- form
the basis of the rest of our analysis.  

We can construct a plot of the information in the CMD by summing the assigned
statistical weight of observed sources within each CMD pixel.
This is shown in the fourth column of Figure~\ref{fig.examplemaps}.
Most of the parallax signal comes from a narrow vertical stripe in the CMD
corresponding to Galactic turnoff stars.
About half the total
comes from sources in the color range $0.6 < BP - RP < 1.4$, and magnitude
range $16 < G < 18$.
We will thus refer to our sample as ``turnoff stars'',
even though we have not actually truncated the sample to include 
only turnoff stars.  
Note that the analogous parallax bias
information in quasars comes from slightly bluer and
  fainter sources (cyan contours in the second row).
The parallax bias information in the RGB sample,
in contrast, comes from much brighter sources (red contours in the third row).
The red clump sample we use is also limited to bright magnitudes,
mainly in the range $12 < G < 14$.

Comparison with the GOG simulation suggests the turnoff stars that
dominate our signal are typically from the 
geometrically-defined \citep[c.f.][]{gilmore83} thick disk component of this model,
with absolute magnitude about 5 and a distance $\tsim 1 \kpc$.
There are typically 30k--1M sources in this
primary signal region within each circular field (or about 4--150 per pixel),
and 300k--5M sources overall (about 40--600 per pixel),
increasing towards the Galactic plane.  The huge variation in source density
will result in strongly varying noise of the bias maps over the sky.

We use the \citet{schlegel98} maps to define spatial pixels that should be excluded from initial analysis.
Our chosen procedure retains most of the sky pixels but cleans the
sample of the worst effects of dust clouds, though this is
judged by visual impressions rather than any formal optimization method.
Specifically, we first perform $1.7\sigma$-clipping on the 
$E(B-V)$ values of the spatial pixels within each circular field.
We use the upper threshold defined in this clipping, or an
$E(B-V)$ of 0.1, whichever is greater, to define a threshold
above which pixels should be excluded.  This method excludes most
of the dense dust clouds, but allows regions with large but
relatively smooth extinction to contribute.  
Sources within these masked pixels are not included in CMD measurements,
and are also excluded during spatial smoothing, while we allow
surrounding pixels to fill in the excluded region.

We also want maps that select or exclude strong artifacts or compact
physical features, so we can single out the more or less typical
systematics behavior in turn.  To do this we smooth the parallax and
delta-parallax maps with a Gaussian smoothing length of $0.4 \degree$, which
reduces the waffle pattern's influence. We next subtract identically smoothed
versions of the corresponding parallax or delta-parallax GOG simulation maps,
which helps flatten large-scale trends in the maps due to
stellar population gradients.
We then subtract off the means of the maps, to give them
zero mean, and combine the parallax maps $\cal{M}_\pi$ and
delta-parallax maps $\cal{M}_{\delta \pi}$ in a
linear combination that emphasizes the parallax maps near the disk and
the delta-parallax maps near the poles, combining the advantages of
each type. (Specifically,
we use the combination $u \, {\cal{M}}_{\pi} + (1-u) \, {\cal{M}}_{\delta\,\pi}$)
with $u = (1 - |b|/(90\degree)^2$.)
Pixels beyond $20 \muas$ from zero (approximately 2$\sigma$)
are then marked as candidate strong artifacts.
We will use these artifact regions in Section~\ref{sec.scars}.
Combining these maps with any pixels previously excluded by our dust maps
yields our ``combined'' exclusion maps, for use in studying the typical
behavior of the parallax bias.

The regular variations on small scales seem connected to the
\gaia\ scanning pattern \citep{lindegren18}.
We checked this using models of the \gaia\ scanning path as a function of time,
at first using a model provided by Sihao Cheng and Sergei Koposov,
\footnote{{\tt github.com/SihaoCheng/Gaia}}
and later confirming with a model provided by the \gaia\
collaboration\footnote{{\tt www.cosmos.esa.int/web/gaia/scanning-law-pointings}}.
We verified in numerous fields with a strong waffle pattern that the streaks
align with at least one scanning path orientation.  In each case
the scan paths fall into two or three groups of nearly equal orientation,
and the waffle pattern reflects this in the presence of two or three
dominant streak patterns.  However, it is not clear to us
how to predict the exact bias pattern from the scanning law,
and we make no direct use of the scanning path in the analysis.

We have constructed median-filtered maps of various other quantities,
including the proper motion, the parallax error, the correlation of parallax
and proper motion,
the astrometric noise, and the RUWE.
We found no simple relationship between these quantities and the parallax variations.  
It seems clear that all maps show regular variations on a similar angular scale,
clearly connected to the \gaia\ scanning pattern,
but not reliably peaking in the same locations or showing the same combination
of wave amplitudes as the parallax bias maps.
In Section~\ref{sec.pm} we will explore connections between the parallax and the
proper motion systematics. For now, we focus on the parallax.
\section{Results}
\label{sec.results}
\subsection{All-sky maps}
\label{sec.maps}
\begin{figure}[t!]
\centerline{\includegraphics[width=142mm]{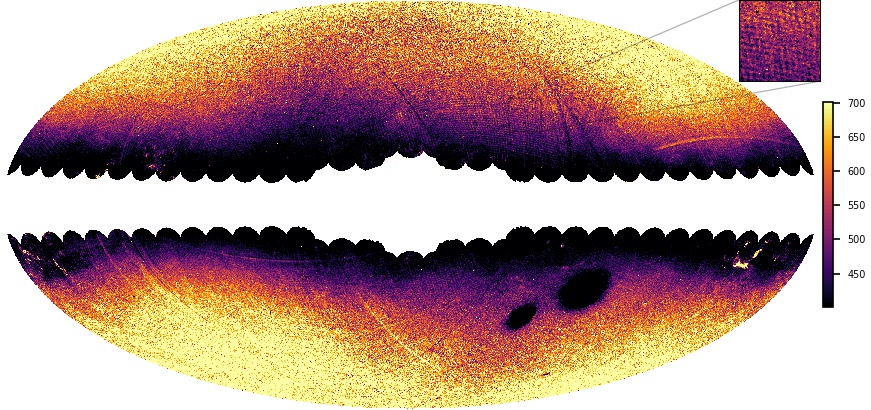}}
\centerline{\includegraphics[width=142mm]{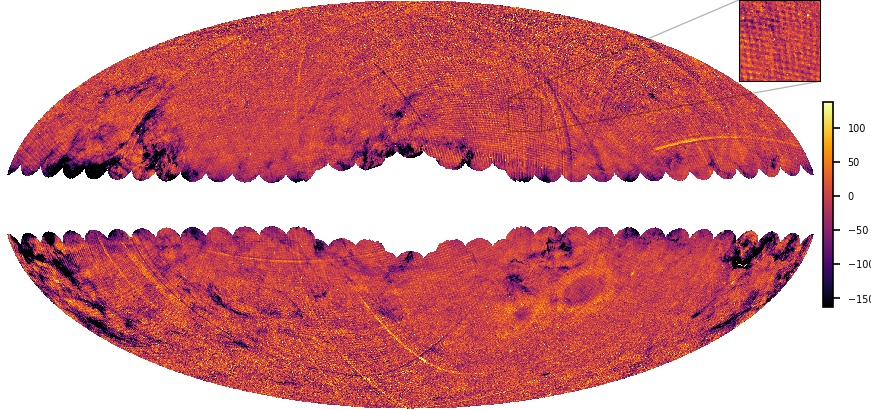}}
\caption{
\label{fig.allskymaps1}
All-sky maps illustrating
the parallax bias from \gaia\ in a Hammer-Aitoff projection.
Top: (weighted average) parallax map.
Bottom: delta-parallax map (average difference from expectation
based on color and magnitude).  
As with other maps in this paper, colorbars show the scale in units of $\muas$.
The adopted range is $\pm 150 \muas$ around the mean value of each plot.
The Galactic Center is in the middle of the maps, with
latitude increasing upwards and longitude to the left.
The Magellanic Clouds are visible as the two rounded distortions
in the southern hemisphere.
}
\end{figure}
\begin{figure}[t!]
\centerline{\includegraphics[width=142mm]{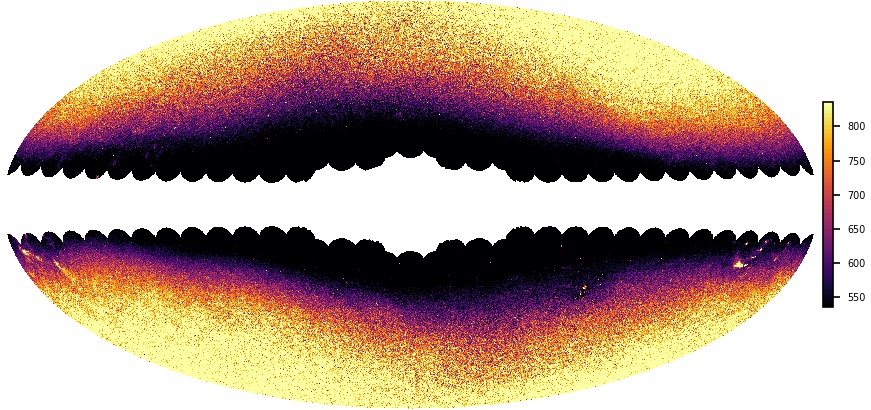}}
\centerline{\includegraphics[width=142mm]{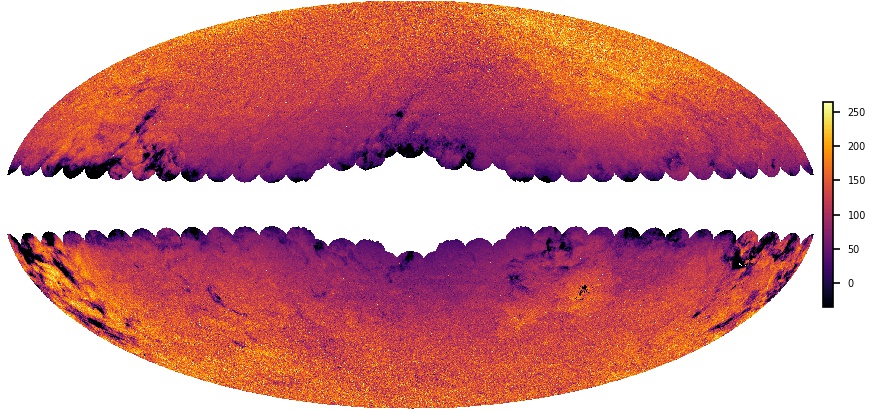}}
\caption{
\label{fig.allskymaps2}
All-sky maps using the GOG simulation.  Orientation
and units are the same as in Figure~\ref{fig.allskymaps1}.
Both maps use \gaia\ as the reference sample to define
weights for averaging.
Top: (weighted average) parallax map.
Bottom: delta-parallax map (difference from expectations).
Here \gaia\ data also defines the expected parallax,
so this map is not flat on large scales unlike the analogous observed map.
}
\end{figure}
\begin{figure}[t!]
\centerline{\includegraphics[width=142mm]{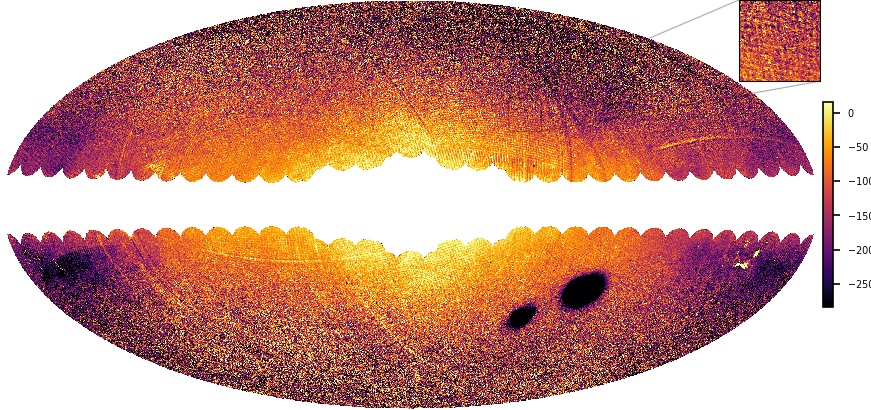}}
\centerline{\includegraphics[width=142mm]{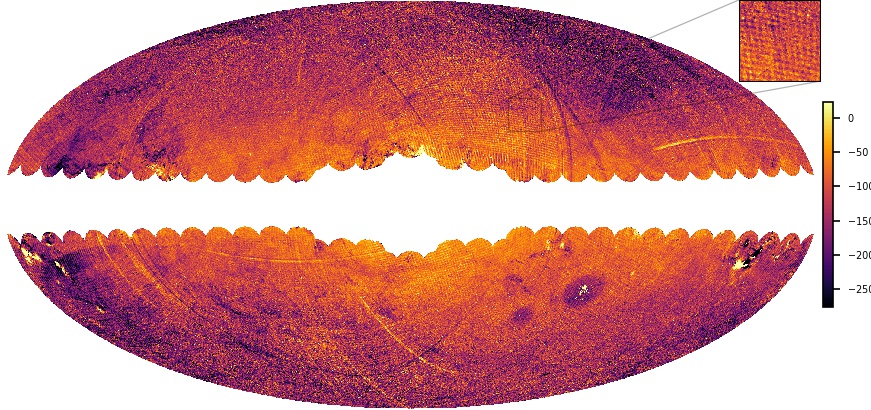}}
\caption{
\label{fig.allskymaps3}
All-sky maps showing \gaia\ after subtracting GOG maps.
The parallax map is on top, and delta-parallax map on the bottom.
Orientation and units are the same as Figure~\ref{fig.allskymaps1}.
}
\end{figure}
The top panel of Figure~\ref{fig.allskymaps1}
shows a map of the weighted mean parallax of the sample stars over the sky.
This is assembled from individual maps of our $6 \degree$-radius fields,
combined into a Hammer-Aitoff projection using the IPAC Mosaic code
\footnote{{\tt github.com/Caltech-IPAC/Montage}}.
The map shows a combination of astrophysical and instrumental effects.
The primary astrophysical effect is a strong gradient with
Galactic latitude.  This occurs because at low latitudes,
disk stars are on average located further away.
The LMC and SMC are also clearly visible as round dark patches 
below and to the right of the Galactic center.
The effects of dust clouds are visible in a few light patches.  

However, various effects in this map appear to be \gaia\ systematics,
due to their regularity and/or distribution on great circles.
Several widely prevalent effects of distinct character can be recognized.
These include the regular waffle pattern visible on small scales,
as well as individual narrow streaks we term ``scars''.
The persistence of these features over scales much larger than our individual fields
shows that our method produces consistent results in different parts of the sky.
There are also asymmetric variations on larger scales which may not be astrophysical,
though this can be probed better with maps we will construct later.

The bottom panel shows the analogous delta-parallax map over the sky.  
Since the average of delta-parallax in each field is (by construction)
very close to zero, the overall map is rather flat.
The effects of dust clouds here show up much more prominently as dark filamentary patches.  
Artifacts due to the circular fields are visible in some regions, especially around the
Magellanic Clouds and Galactic plane.
The waffle pattern and scars are much more obvious, due to the reduced noise
with this method.

Figure~\ref{fig.allskymaps2} shows the analogous maps constructed
from the GOG simulation.  
Note the simulation includes detailed modeling of Galactic dust clouds,
as well as models of the Magellanic Clouds, though 
we cannot rely on complete accuracy in treating these complex features.
Most astrophysical effects seen in the \gaia\ maps are clearly evident in the simulated maps
as well, while the instrumental systematics have vanished.
Some artifacts due to the individual circular fields are apparent.  
Recall that the delta-parallax map of the simulation is constructed
using the observed \gaia\ sample to set expectations for the parallax
at a given CMD location.
Hence this map shows much more large-scale variation than the
observed delta-parallax map, which compares the \gaia\ data to itself.

Figure~\ref{fig.allskymaps3} shows the results of subtracting the simulated maps
from the observed maps.
This roughly corrects for the various astrophysical effects, though the correction is
clearly not perfect --- note the strong dependence on Galactic latitude,
especially in the parallax map.
The various types of \gaia\ systematics are now more clearly evident.  
Note in particular the large area of low parallax in the northern Galactic hemisphere,
  which has roughly straight boundaries meeting at a corner at $l \sim 270 \degree$, $b \sim 30\degree$.   
The parallax map is noisier at high latitudes, due to the lower stellar density there,
  but at low latitudes displays the systematic variations more clearly 
  because of its lower sensitivity to dust effects.
See in particular the strong edge at $l \sim 150 \degree$, $b \sim -20 \degree$.

\subsection{Waffle pattern}
\label{sec.waffle}
The most consistent systematic in the parallax maps is the
quasi-regularly-repeating signal we term the ``waffle pattern'',
previously depicted using LMC and SMC stars in \citet{lindegren18} and \citet{helmi18}.
As seen in those depictions and in Figure~\ref{fig.examplemaps}, this
normally consists of two or three superposed wave-like features, giving
the appearance of square or triangular cells.  Some of
these wave features are concentrated in narrow streaks of either sign,
while others are more sinusoidal.  The character of this pattern
changes from region to region, sometimes slowly and sometimes abruptly.
To analyze this pattern, we apply the ``combined'' exclusion maps
  described above, which are intended to avoid dust features
  and strong artifacts like the scars.
  
\begin{figure}[t!]
\centerline{\includegraphics[width=134mm]{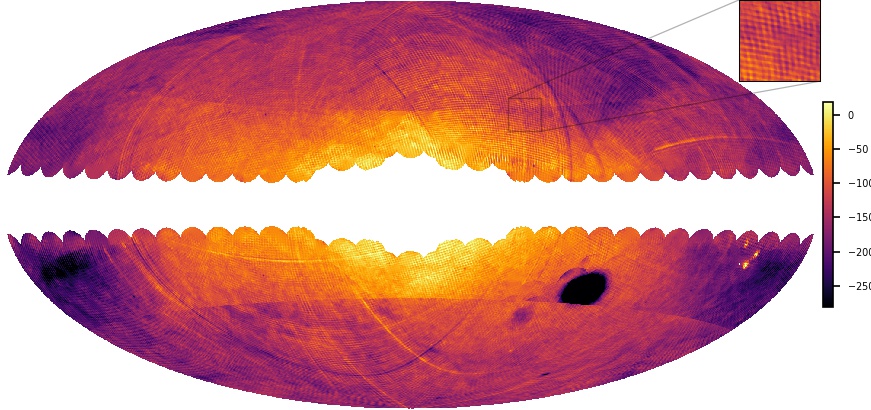}}
\caption{
\label{fig.allskymaps_smoothed}
Wiener-smoothed map of the parallax bias over the sky (see text for details).
Orientation and scale are as in Figure~\ref{fig.allskymaps1}.
The mean-parallax method is used
for $|b| < 38\degree$ and delta-parallax map for higher latitudes, which
causes the visible mismatch at those lines.
To correct for genuine parallax variation, we smooth the GOG simulation maps on a scale of 
$0.5\degree$ and subtract these from the observed bias maps before display.
Overall calibration issues remain to be corrected in a later step.
}
\end{figure}

The small angular scale of the waffle pattern makes it difficult to
map precisely, especially in regions of low stellar density.
The raw pixel values from our initial maps are fairly noisy,
especially at high Galactic latitude.
Simple Gaussian smoothing, if using a smoothing length large enough 
to reduce the noise, tends to blur the waffle pattern too much to be useful.
The most effective smoothing method we have found is a Wiener-like filter,
  where we specify models for the signal and noise component variance in Fourier space,
  and explicitly include wave-like features in the signal model.
To construct the component models used in this filtering method,
we first Fourier-transform all the usable maps, compute the 
  squared amplitudes of the modes, and average over the maps.
The resulting 2-d spectrum shows a central peak in power, which we fit
  with a power-law component, and also shows a ring at wavelength $\tsim 1 \degree$
  resulting from the waffle-pattern modes.  
To smooth the individual maps, we first measure the noise level in the map from the
  mean amplitude of high-frequency modes.
We also identify the strongest peaks in
Fourier space with wavelength around $1 \degree$,
  incorporate several harmonics of these peaks, 
  and add the combination as additional components
  of the signal model to the previously fitted central component.
These signal and noise components are then used to construct
the Wiener filter we apply to the observed power spectrum.
It is important to note we are not fitting the precise Fourier
transform amplitudes and phases with a model in the peak-selection step,
which would place high demands on the model accuracy,
but rather constructing a kind of pass-through filter
in order to keep the waffle-pattern modes while suppressing the noise.
The result of this smoothing procedure over the whole sky
is shown in Figure~\ref{fig.allskymaps_smoothed}.
(Here we have switched from parallax to delta-parallax maps at intermediate
latitude, to combine the better high-latitude precision of the latter with the
smaller dust-induced artifacts of the former.)
If we designate only the peaks as signal for the Wiener filter,
we obtain a differently filtered
map that displays only the waffle pattern and leaves out larger-scale
features, which will be useful later.

The waffle pattern is ubiquitous on the sky, as seen in
Figure~\ref{fig.allskymaps3} and Figure~\ref{fig.allskymaps_smoothed}.
However, its amplitude appears to vary with sky position.
Compare for example the regions just north of the Galactic plane
at longitudes $\tsim 60 \degree$ to the left and right of the Galactic center.
To quantify this variation, we use the
``combined'' mask maps to avoid pixels heavily 
contaminated by dust, discrete stellar systems, or strong artifacts.
We discard entirely any fields where more than 35\% of the pixels are masked by our
combined exclusion maps, leaving a ``clean field'' sample of 492 maps (out of 590)
for the small-scale analysis. 
We then measure the local amplitude of the waffle pattern within our
circular fields, using two different methods.
In the first, we simply took the rms value of the Wiener-smoothed delta-parallax maps, 
masking out dust clouds and ``scars'' as before.
Estimates based on smoothed maps have the potential to oversmooth
the waffle pattern and underestimate the amplitude
(as we found with some of our initial smoothing attempts).
Therefore, we also
compute the correlation of pixels with their immediately adjacent map pixels
in the clean field sample,
and estimate the waffle pattern amplitude as the square root of this correlation value.
This method should overestimate the strength of the waffle pattern, if anything,
due to the contribution of pixel noise to the variance.
The two measures are reasonably consistent for individual fields,
and have similar overall distributions.
For the smoothed-map method, we find a waffle-pattern rms amplitude of
$15 \muas$ with dispersion $3 \muas$ (or 20\%).
The central 50\% of the fields have amplitudes
in the range $12$ to $16 \muas$, and 90\% are within the range $10$ to $21 \muas$.
For the noisier correlation method, we find an rms amplitude of
$19 \muas$ with dispersion $14 \muas$.

\citet{lindegren18} previously estimated the systematic error of the parallax
using quasars.  They found the correlation function could be represented as
a long-distance term plus an oscillatory excess induced by the small-scale pattern.
Subtracting the long-distance term, their zero-separation correlation equates to
an rms bias variation $\approx 40 \muas$.
(This value was later adopted by \citealp{vasiliev19} in their formal treatment
of parallax systematics.)
Our typical rms amplitude of $15 \muas$ is only $40\%$ of this value derived from the quasars.
The amplitudes of the parallax variations plotted for the LMC in \citet{helmi18},
  however, seem more consistent with our lower value.
  As a consistency check, we constructed a map of the (unweighted) mean parallax
  within a $3 \degree$ radius around the LMC.
This map has an rms variation of $18 \muas$, in good agreement with our typical
value and again much lower than the value derived by \citet{lindegren18}.  
It should be noted that the statistical errors on the quasar correlation function
  at small separation are quite large, due to the rarity of close pairs of quasars,
  and we suspect it to be statistically consistent with our result.  
This difference could also be influenced by a varying response
of different types of sources to the waffle pattern, and we now
check for such a variation.  
  
\begin{figure}[t!]
\centerline{
\includegraphics[width=65mm]{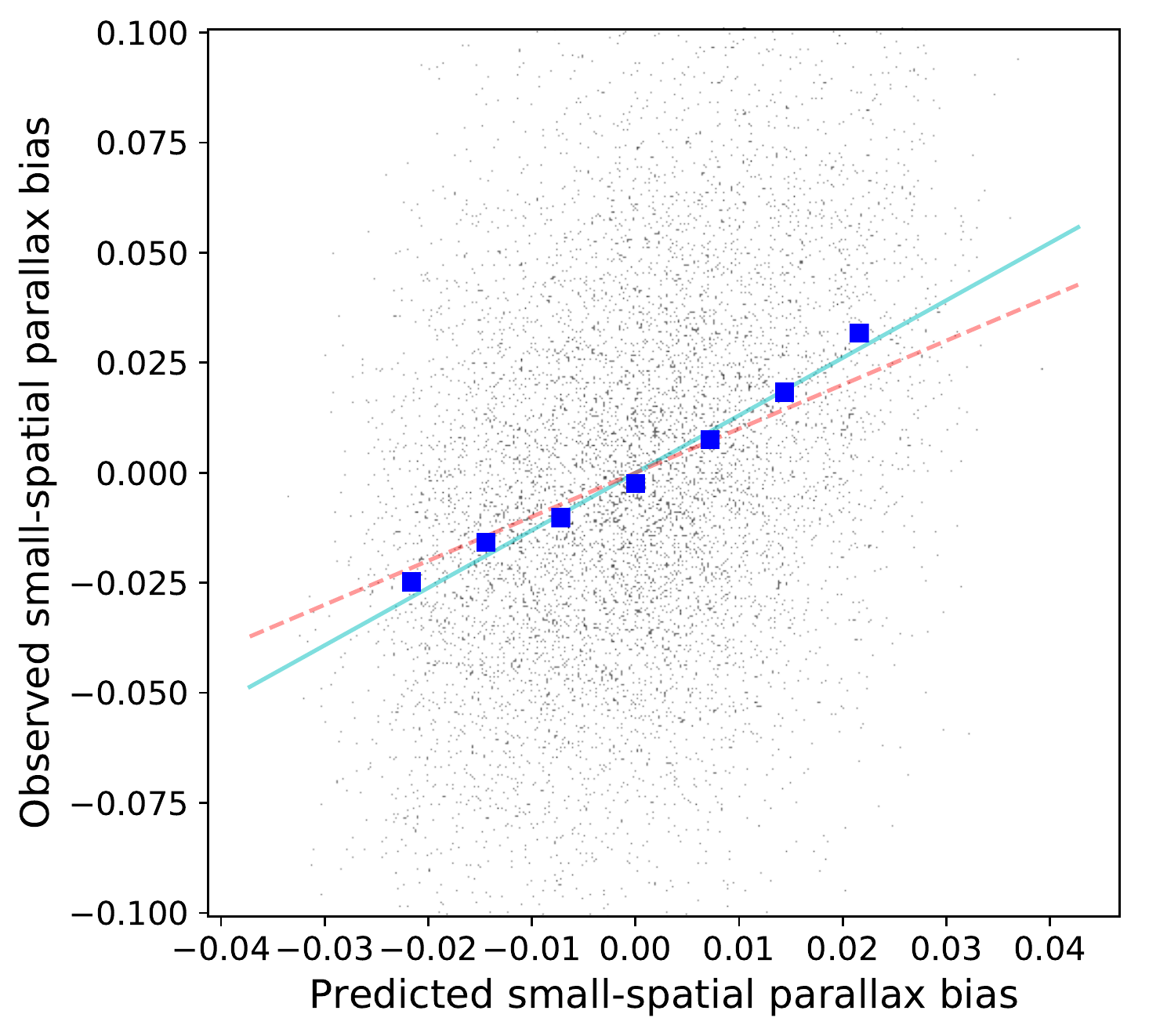}
\includegraphics[width=75mm]{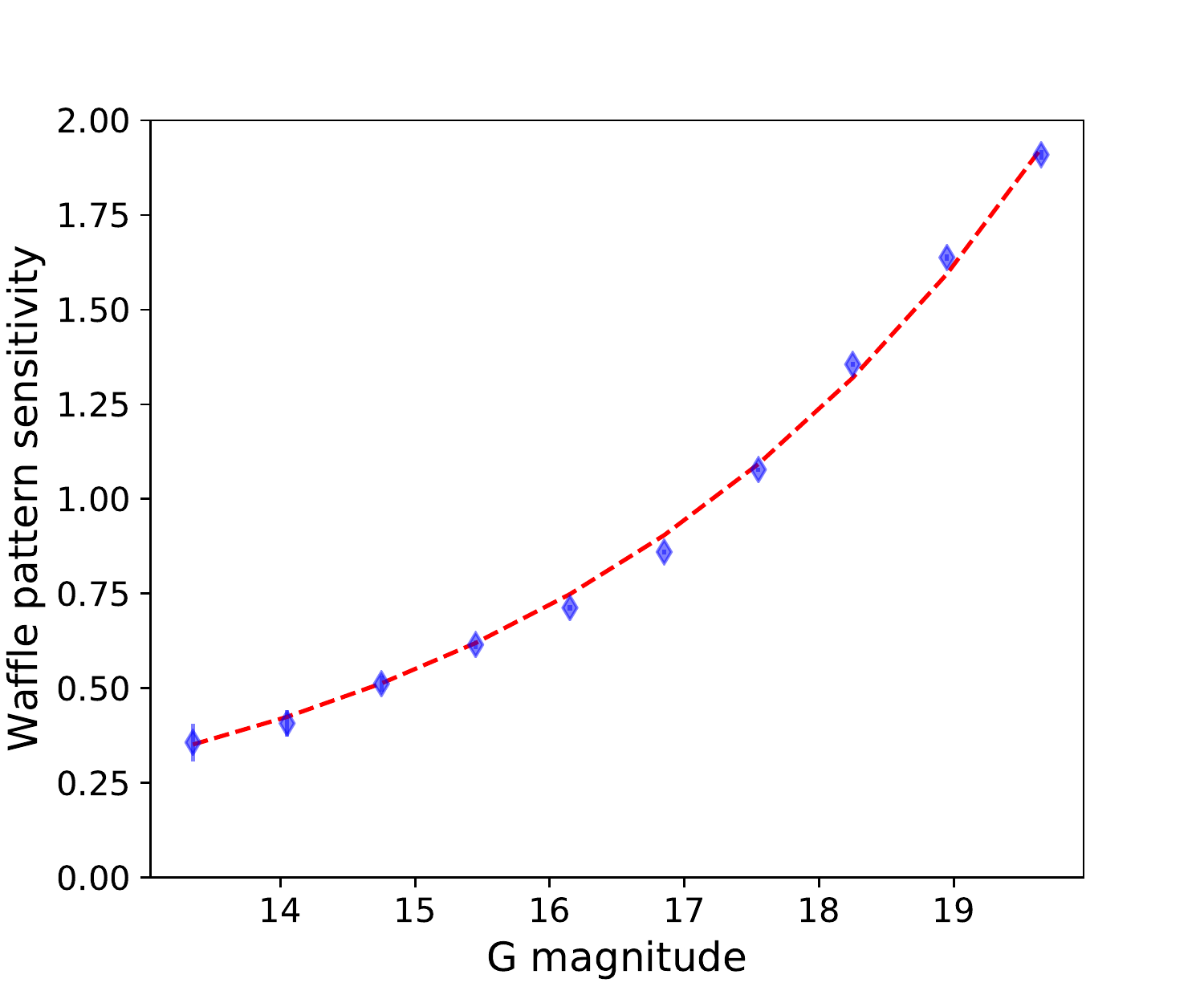}
}
\caption{
\label{fig.magdependence}
Left: waffle pattern sensitivity estimation.
Plot shows delta-parallax map pixels
made from sources in one of our circular fields
in the magnitude bin $17.2 < G < 17.9$,
versus the delta-parallax value in the same spatial pixel
using a map made from the sources outside this magnitude bin
(see details in text).
Blue squares show binned averages.
A linear fit of observed versus predicted values is shown by the blue solid line.
The red dashed line shows the 1:1 relationship.
We denote the slope of this line as $f_w$, the waffle pattern sensitivity.
Right: waffle pattern sensitivity to magnitude.  Blue points show the
$f_w$ slopes obtained for entire sample of magnitude bins, averaged
over spatial fields, as a function of magnitude.  The red dotted line
is the fit given in the text.
}
\end{figure}

Restricting the parallax bias maps to sources in different magnitude
ranges, we observe that the apparent strength of the waffle pattern
increases towards faint magnitudes.  We quantify this on a subset of
fields with high stellar density, where the sampling noise will be
minimal even after subdividing the data.  We first split the source
magnitude range $13 < G < 20$ into 10 equal bins.  For each bin, we construct 
an unsmoothed delta-parallax ``response'' map using only sources within the bin,
and a delta-parallax ``signal'' map using only sources outside the bin.
To suppress noise in the signal maps, we Wiener-smooth them in the manner
described above, except here we designate only the waffle component as a signal
component.  This minimizes larger-scale variations from population gradients and dust.
We then plot the in-bin response value versus the out-of-bin pattern value,
excluding pixels flagged by our combined-exclusion maps,
and fit a line to this relationship,
as in the left panel of Figure~\ref{fig.magdependence}.
Repeating the process for each bin, we obtain an estimate for the sensitivity
$f_w$ to the weighted-average version of the waffle pattern
as a function of source magnitude, in each field.
Averaging these slopes yields a measure of $f_w$ over the whole sample
(right panel of Figure~\ref{fig.magdependence}).
This shows a steep, steady rise toward faint magnitudes, increasing by
over a factor of 6 over the magnitude range $13 < G < 20$.
A reasonable fit over this range is $f_{w} = \exp[0.27 \, (G - 17.2)]$.
The results are not changed much if we fit the slopes over the entire sample
of fields, rather than fitting field by field and then averaging the slopes.
We also test for a relationship versus color.  The sensitivity as a function
of color does show some variation, but it is hard to trust the results due to
the high concentration of information at the color of turnoff stars.

\begin{figure}[t!]
\includegraphics[width=44mm]{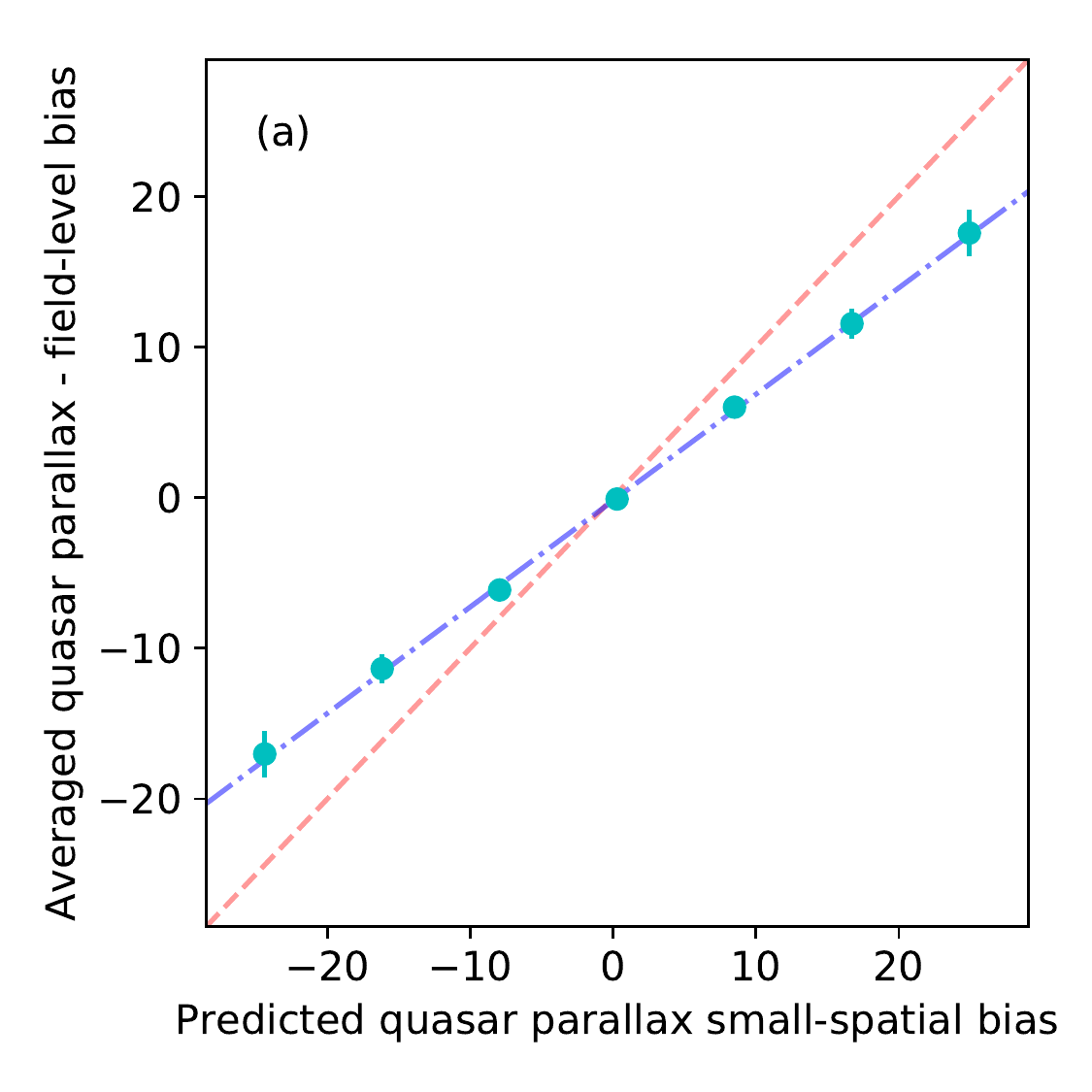}
\includegraphics[width=44mm]{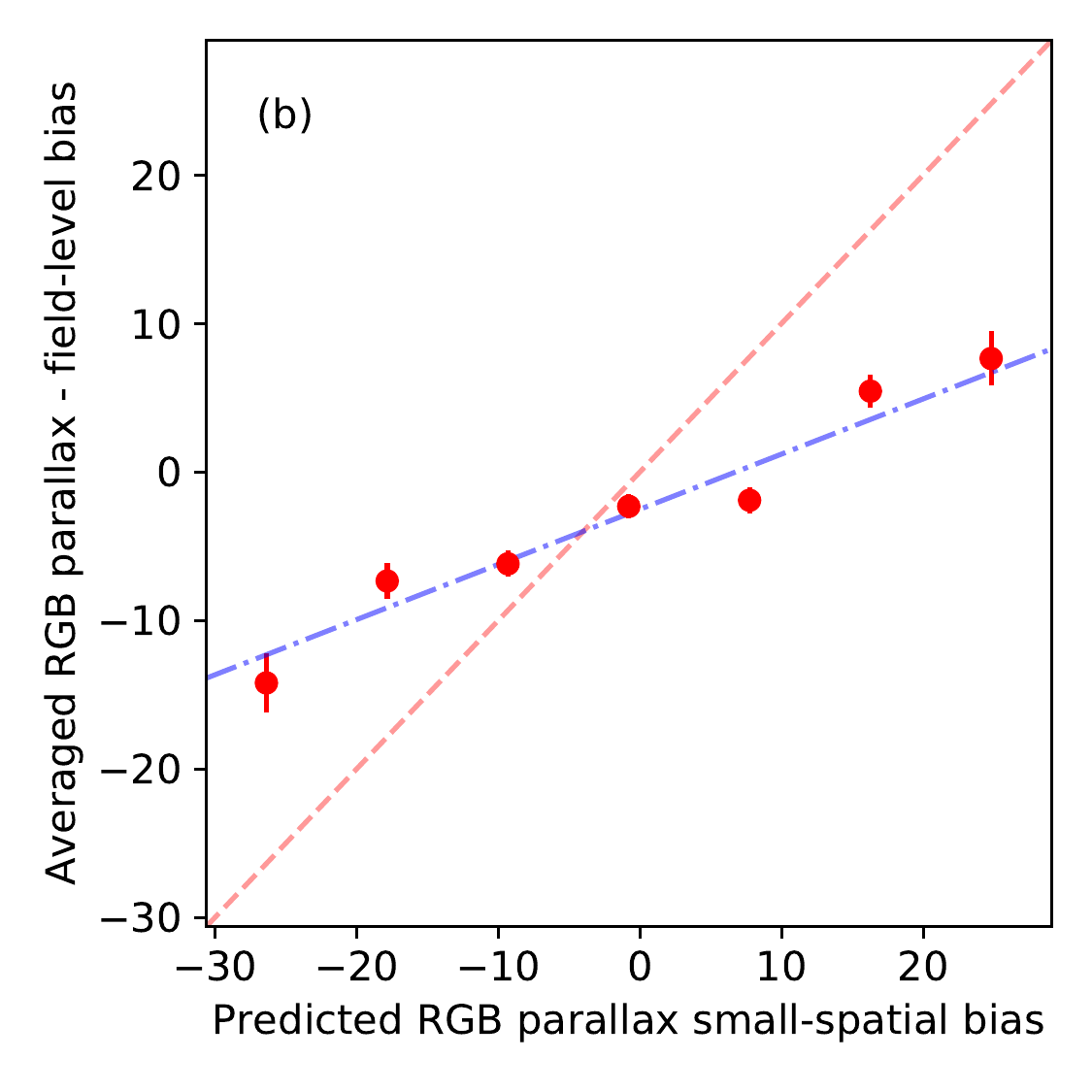}
\includegraphics[width=44mm]{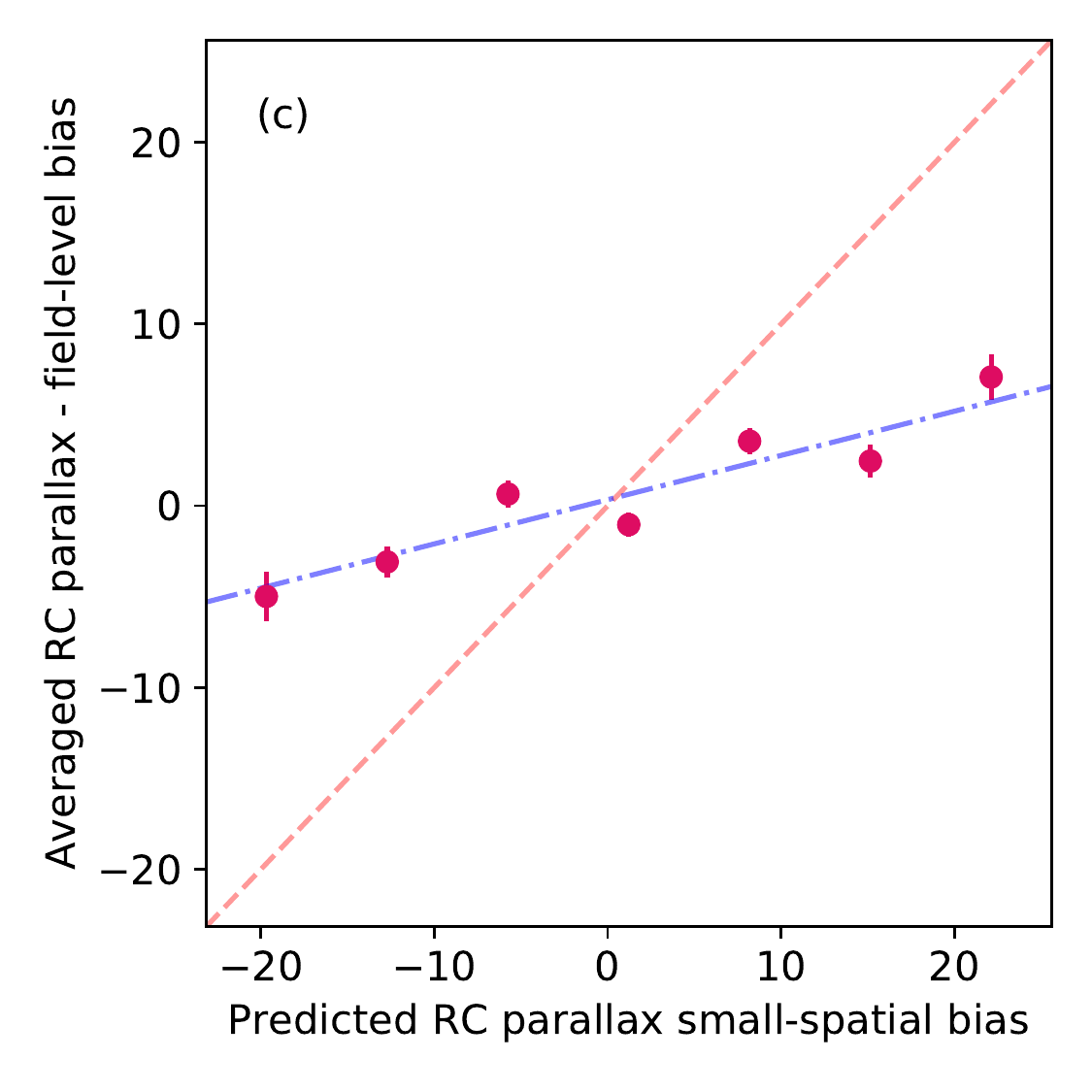}
\includegraphics[width=44mm]{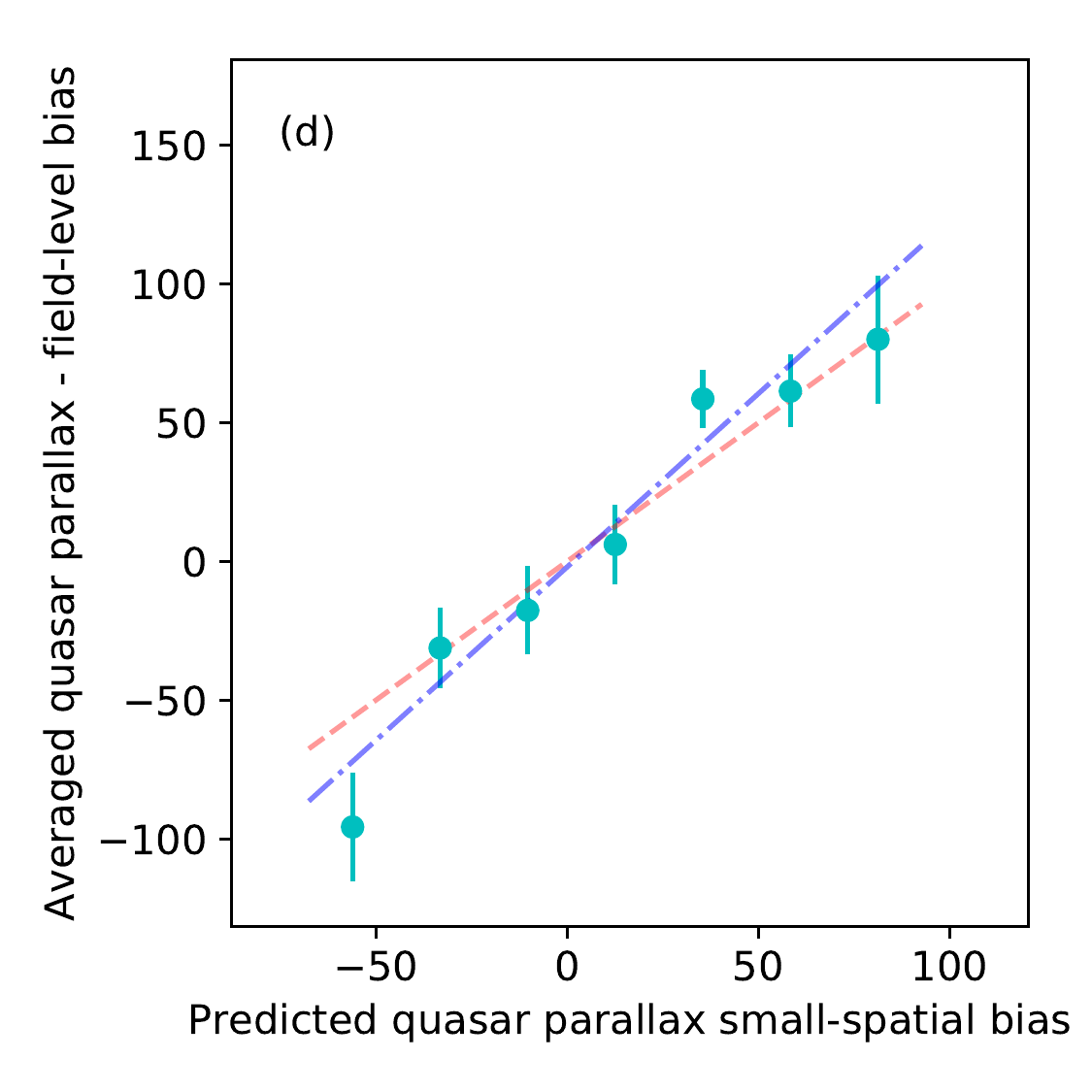}
\caption{
\label{fig.smsclsrccorr}
Panel (a): correlation of quasars with the waffle pattern.
Red line indicates the 1:1 relationship.
Dashed line is the best fit.
Panel (b): correlation of RGB stars with the waffle pattern.
Lines have the same meaning as in the previous panel.
Panel (c): correlation of Kepler-field red clump stars with the waffle pattern.
In each of these cases, the average of quasar parallaxes within
the field has been subtracted from the source parallax.
Panel (d): correlation of quasars with scar features.
Note the much broader data range of this plot, and the slope close to unity.
}
\end{figure}

While the waffle pattern seems clear enough within the turnoff stars,
it is worth showing that it extends to other types of sources,
both to test across a wider range of color and magnitude 
and to verify it is not somehow artificially produced or amplified by our analysis procedures.
We first check versus the quasars in the \citet{secrest15} catalog,
which are on average somewhat fainter and bluer
than the stars dominating the turnoff sample (see their information
contours in Figure~\ref{fig.examplemaps}).
We use the Wiener-smoothed delta-parallax maps as a prediction of the local bias.
We sample these maps at the quasar locations,
subtract the average quasar parallax in the given field
to cancel out large-scale fluctuations,
and compare the resulting values to the measured quasar parallax.
For this correlation test, it is important
that we have excluded sources in the quasar 
catalog from the stellar sample, to avoid biasing the maps.
Binning and averaging the source parallax bias, using
the estimated parallax errors on individual quasars to weight the average,
we find a smooth
linear trend of mean observed versus predicted parallax bias (panel (a) of Figure~\ref{fig.smsclsrccorr}),
showing the quasars definitely respond to the waffle pattern measured in the turnoff stars.
However, the slope is only $0.71 \pm 0.01$, significantly less than unity.
If the measured turnoff star contains both true waffle-pattern
signal with variance $V(s)$ and
noise with variance $V(n)$, and the quasars depend on the waffle pattern
with proportionality constant $f_w$, 
we expect a slope of $f_w V(s) / [V(s)+V(n)]$.
Samples at higher latitude, and/or maps that are more careful about excluding
dust-contaminated regions, yield slopes $\tsim 0.8$ instead, closer to unity.
This might suggest the predicted bias maps include contributions from purely
astrophysical effects, of amplitude roughly half that of the true \gaia\ bias signal
with variations due to location.  
Still, it is reassuring that we find the quasars correlate so strongly 
with the waffle pattern indicated by the Galactic stars,
as it validates this systematic using a sample of sources
that have a reliable true parallax of zero and lack the
dust-induced biases present in the turnoff sample.  

Stars with parallax values estimated from spectroscopy
present an interesting test of the waffle pattern 
for sources much brighter than our turnoff stars
(see the contours in the last panel of Figure~\ref{fig.examplemaps}).
We use the RGB sample of \citet{hogg19} in a similar manner as for the
quasars, though here we cannot assume the parallax to be zero
and instead measure the bias via the difference of \gaia\ parallax
from the value inferred from spectroscopy.
We again weight by estimated error on the bias, 
derived from the estimated spectroscopic error and the \gaia\ statistical
errors added in quadrature.
Presumably, any overall biases in the spectroscopic estimate
will vary on a much larger angular scale than the waffle pattern.
In the overall sample, we again find a clear trend with predicted bias
(panel (b) of Figure~\ref{fig.smsclsrccorr}), 
but with a slope of $0.37 \pm 0.03$, lower than that found for the quasars.
This indicates a weaker waffle pattern in the RGB than in turnoff stars or quasars.
This seems consistent with the magnitude trend we derived for the turnoff stars.

Finally, we perform a similar exercise on the red clump stars
in the Kepler field from \citet{hall19}.
We estimate the true source parallaxes from their observed $K$-band magnitudes,
which \citet{hall19} argue have a small intrinsic dispersion.
For these stars we again find
a good trend of parallax bias with waffle pattern value
(panel (c) of Figure~\ref{fig.smsclsrccorr}),
with a slope $0.23 \pm 0.027$, markedly lower than for the quasars.
These stars are all relatively bright.  For sources with $G > 13$,
we derive a higher slope of $0.32 \pm 0.026$, similar to the bright RGB stars,
while for sources $G < 13$, we find a very weak slope of
$0.098 \pm 0.042$.  
Taken together, these results on RGB and RC stars and quasars confirm
that the waffle pattern
extends beyond turnoff stars to other regions of the CMD.
They also add to the evidence that the strength of the waffle pattern
varies with magnitude.
We conclude that we should {\it not} rely on estimates of the small-scale 
systematics using quasars or LMC stars, or the more general measurements here,
to set expectations on small-scale systematic errors for the brighter stars
with $G \lesssim 14$.

\subsection{Scars}
\label{sec.scars}
Earlier we described our automatic selection of candidate strong artifacts
in the maps.  We now examine the strong linear features
we have termed ``scars'', easily visible crisscrossing the sky
in Figure~\ref{fig.allskymaps3}.
We select fields with obvious scar regions through visual inspection
of the maps.  We next cluster the candidate pixels in each of these fields
using a friends-of-friends
algorithm, and retain only the largest group in each field to exclude noise
peaks or isolated non-scar features.  This automated procedure is usually
successful in detecting segments of scars, but we re-inspect the maps to
exclude those where it was not. This yields a sample of scar region pixels.  

The strong scar artifacts occupy only a small fraction of the sky.
While they roughly follow great circles,
they do not necessarily align with the orientations of the various
  strong waves of the waffle pattern.
The scars are superficially similar in morphology to narrow tidal
  streams from globular clusters, though it seems implausible that diffuse
  tidal features could leave visible imprints in the mean parallax.
To be on the safe side, we use the
  {\tt galstreams}\footnote{{\tt github.com/cmateu/galstreams}}
  package to check that the apparent scar features are not coincident
  with any of the tidal streams catalogued there.

The parallax bias of the pixels we identify as scar features,
  relative to the average within their fields,
  have dispersions of $40 \muas$ with tails out to $\pm 100 \muas$ or so.
These are significantly larger offsets than the typical strength of
  the widespread waffle pattern, as shown above.
Furthermore, in most cases the linear scars are wider than the stripes
  forming the waffle pattern.  
In some areas, positive and negative scars run parallel to each other.
It is perhaps fortunate that we do not see any strong scars within
the Kepler field, where the parallax bias has been studied intensively.

We analyze the dependence of quasar parallax on scar intensity
using a method identical to our analysis of the waffle pattern.  
The result is shown in panel (d) of Figure~\ref{fig.smsclsrccorr}.
The fitted slope yields a sensitivity of $1.19 \pm 0.12$, close to unity.
Clearly, the quasars feel the systematic scar artifacts 
just as strongly as the Galactic turnoff stars do.
The overlap of the bright RGB sample with the scar features is not very
extensive, so we will not show a plot from the analogous exercise,
although we note the correlation is positive.
\subsection{Large-scale pattern}
\label{sec.largescale}
We now examine the large-scale variations in the parallax bias seen in
the previous all-sky maps.
These have an uneven topography including both gradual variations over tens of degrees
and sudden jumps over a distance of a degree or so.
Some of these variations may be related to astrophysical effects, like
  dust or mismatches between the real Galaxy and the GOG model.
However, some of these variations seem to be \gaia\ artifacts due to
their regular boundaries, for example the large area of low parallax in the Northern
  Galactic hemisphere.  
These features can also be seen, though less distinctly,
in plots using quasars from the \gaia\ release papers
\citep{lindegren18, mignard18, arenou18}
and later work \citep{khan19,chan20}.
This encourages us to try to use quasars to calibrate the turnoff star
parallax bias on large scales.

We use the simulation-corrected parallax and delta-parallax maps 
  to estimate the parallax bias in each field, averaging over unmasked pixels.
  We construct a similar average from the quasars within the field,
  and subtract the mean bias $-30 \muas$ \citep[c.f.][]{lindegren18}
  from each value to leave only the local offset from the mean.
The raw averages from turnoff stars show only a weak correlation with the quasars,
  indicating significant offsets between the two measures that exceed the
  formal statistical errors in the quasar values.  
However, there are strong trends of the offset with both latitude and longitude, 
  as indeed might be expected from Figure~\ref{fig.allskymaps3}.
We assume symmetry of the Galactic disk population in longitude and latitude,
in which case the calibration should depend on powers of $\cos l$ and $\cos b$.
For both parallax and delta-parallax maps, 
we fit a second-order polynomial in these quantities to the difference
  between turnoff and quasar parallaxes.
We then shift the bias value in all turnoff map pixels by this smooth fit.
The resulting averages for the delta-parallax maps agree somewhat
better with the quasar field averages than the parallax maps do,
suggesting the smaller astrophysically-induced gradients in the former method
(see Figure~\ref{fig.allskymaps3}) are easier to correct.

\begin{figure}[t!]
\centerline{\includegraphics[width=142mm]{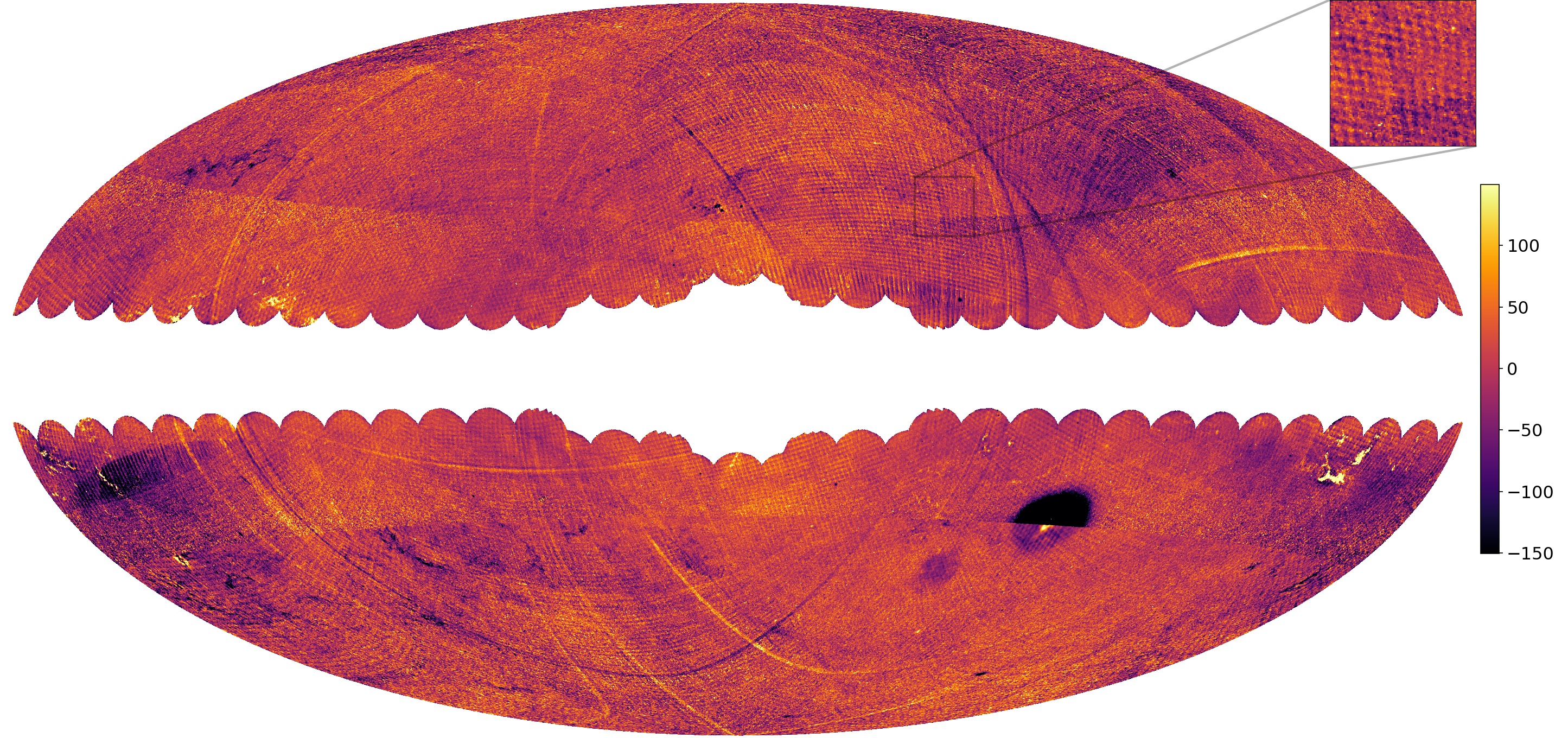}}
\caption{
\label{fig.allskymaps_flattened}
Map of parallax bias over sky calibrated on large scales,
with mean-parallax method used
for $|b| < 38\degree$ and delta-parallax map for higher latitudes.
For each method the simulation map is subtracted,
then the result has been flattened using a large-scale calibration
versus quasars.
A mismatch is usually visible at the join lines, while other sharp
edges are \gaia\ parallax artifacts.
Orientation and units are as in Figure~\ref{fig.allskymaps1}.
}
\end{figure}

In Figure~\ref{fig.allskymaps_flattened} we show the results of this
large-scale calibration over the sky.
Here we use the unsmoothed simulation-subtracted maps of Figure~\ref{fig.allskymaps3},
and simply shift each field map by the large-scale calibration estimate.
As in Figure~\ref{fig.allskymaps_smoothed}, we
use the parallax maps for $|b| < 38\degree$ and the delta-parallax maps
for $|b| > 38\degree$.
(This specific transition is simply due to our visual impression
that a choice between $|b|=35\degree$ and $|b|=40\degree$ gave the
least prominent artifacts in the combined map.)
While this map still contains astrophysical effects
of various sorts, it nevertheless depicts the \gaia\ parallax bias
in considerable detail over most of the sky.
The large-scale contrasts in the quadrant
$180\degree < l < 360\degree$, $b > 0\degree$
now stand out especially well compared to previous maps,
as do the negatively biased features
just to the south of the Galactic plane around anticenter.
The previous all-sky maps have all probed the spatial variation in
the parallax bias in some way.  However, Figure~\ref{fig.allskymaps_flattened}
is the first of our maps that can reasonably be said to represent
the parallax bias itself.

\begin{figure}[t!]
\centerline{\includegraphics[width=160mm]{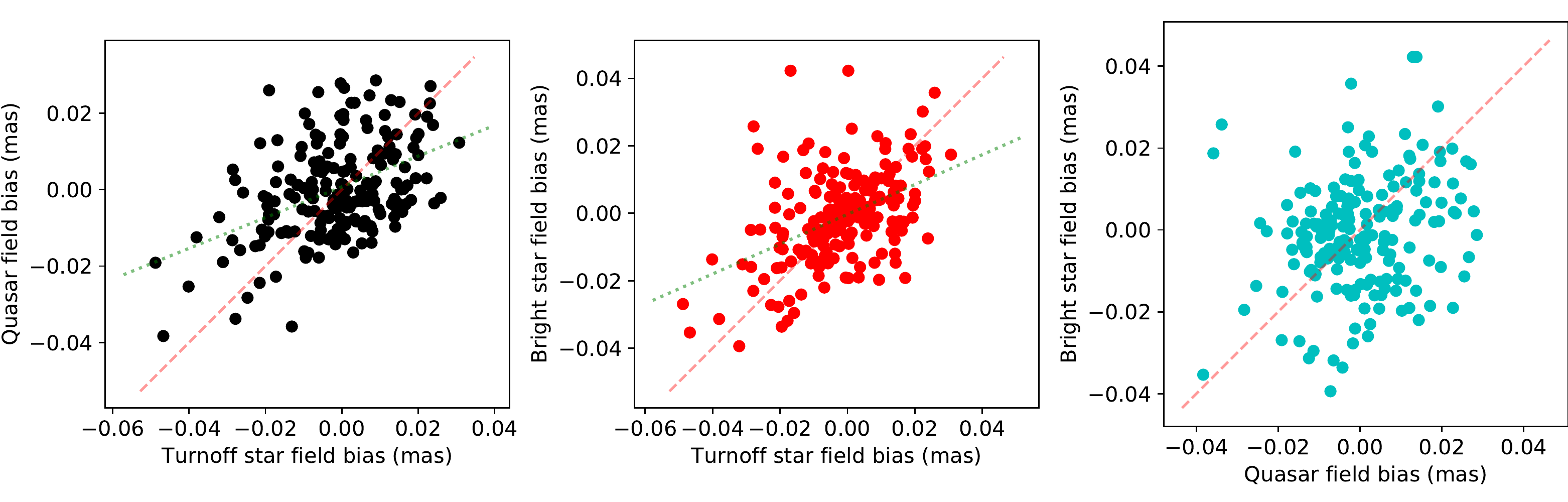}}
\caption{
\label{fig.fieldavg}
Parallax bias averaged over $6 \degree$-radius fields,
compared for different source types.
Left: quasars versus turnoff stars.
Center: bright RGB stars versus turnoff stars
Right: bright RGB stars versus quasars.
The sample of fields is restricted to those in the clean turnoff star sample
and with sufficient RGB stars.
Diagonal lines show the 1:1 relationship. Dotted lines in the first two
panels show the line about which the vertical dispersion is minimized,
with a slope of $\approx 0.4$ in each case.
}
\end{figure}

In Figure~\ref{fig.fieldavg}, we show the field-averaged parallax bias
  estimated from turnoff stars (using the delta-parallax method),
  quasars, and RGB stars.
Here we only display fields in our ``clean'' sample
  which have at least 10 RGB sources, to reduce scatter in the points from 
  large statistical errors in the RGB or offsets from dust clouds.
We can see a clear relationship between these data sources,
though with substantial dispersion in each plot.
Of course, we calibrated turnoff stars to the quasars with a low-order fit,
so we already expect some correlation between those two source types.
The correlation coeffient between the turnoff star and quasar values is
0.46, with a statistical significance of less than $10^{-10}$ chance occurence.
The correlation of RGB values with turnoff star values is
(partly by luck) also 0.46, with a similar statistical significance.
In fact, this is better than the correlation between RGB stars and quasars,
which is only 0.23 with a statistical significance of 0.002.
This implies the bright RGB stars are sensitive
to the large-scale parallax bias seen in the turnoff stars and quasars.
It also suggests our estimate of the large-scale field from turnoff
stars improves on the estimates from quasars, presumably because the
remaining systematic errors in the turnoff stars are smaller than the
large statistical noise in the quasar estimates.

The overall variation in the quasar large-scale bias estimate,
the dispersion in the vertical direction in the first panel of
Figure~\ref{fig.fieldavg}, is about $13 \muas$,
nearly equal to the horizontal dispersion representing the
turnoff star bias estimate.
The RGB dispersion in the second panel is about $14 \muas$,
with a non-Gaussian appearance
possibly due to varying RGB sample size within the fields.
Given the statistical noise in the quasar values,
we estimate the unmodeled (intrinsic) variation is about $10 \muas$,
while the same argument applied to the RGB stars
would imply a similar estimate of $12 \muas$.
Adopting the turnoff star value as a ``corrected'' estimate of the large-scale field
would amount to adopting the diagonal lines in the figure.
Using the vertical dispersion about this line actually increases the value
for both panels, to $14 \muas$ dispersion for quasars and $15 \muas$ for RGB stars.
The line that minimizes the intrinsic dispersion has a slope of roughly 0.4 in both cases.
This probably points to significant noise contamination in the turnoff
star values, of comparable amount to the real signal.
This is also suggested by the fairly large differences between the
parallax and delta-parallax field averages that remain after the flattening procedure,
resulting in the mismatch at the join line in Figure~\ref{fig.allskymaps_flattened}.  
It might be possible to improve the large-scale bias estimators by 
including the locally averaged quasar and/or RGB 
star bias estimators as additional data.
Though this may degrade the high spatial resolution given by the turnoff stars,
it could result in a higher-precision bias estimate over large spatial scales.

We note that unlike the small-scale waffle pattern, the dependence of
the large-scale bias field on source magnitude is poorly constrained
at present.  The similarity of the optimal slopes obtained versus quasars and
RGB stars, which have very different magnitudes, does suggest a weaker dependence
of the large-scale field on magnitude than shown by the waffle pattern.
This could be tested eventually using deeper samples of stars
with independent high-precision distance estimates.

\begin{figure}[t!]
\centerline{\includegraphics[width=134mm, trim={15 10 0 10}, clip=True]{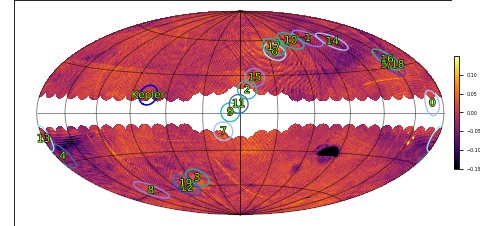}}
  \caption{
  \label{fig.kepler}
  The location of the K2 Campaign (from 0 to 19) and Kepler fields superposed on the delta-parallax map from Figure~\ref{fig.allskymaps_flattened}.  
}
\end{figure}

Intensive asteroseismic studies have been performed in the Kepler and K2 fields,
and parallax estimates are byproducts of these studies \citep{zinn19,khan19}.
In Figure~\ref{fig.kepler} we plot the locations of these fields on the map
from Figure~\ref{fig.allskymaps_flattened}.
We see the Kepler field is in a region of average parallax bias, whereas some
of the K2 fields are in regions of higher or lower bias, differing by about $75 \muas$
from minimum to maximum.
Also, in some of these fields the bias level varies by as much as $150 \muas$ over the field.
Given that we found bright RGB stars clearly respond to the large-scale field, 
the large-scale variations are worth taking into account when interpreting results based
on these fields. (We note \citealp{khan19} previously used the parallax bias
shown by quasars to inform their asteroseismic samples in three of these fields.)

As noted earlier, the \gaia\ parallaxes estimated for globular clusters by
\citet{helmi18} are thought to be unreliable in comparison to astrophysical
estimates, due to systematic errors rather than the tiny statistical errors.
We attempted to apply our parallax bias maps to a sample
of globular clusters to correct the \gaia\ results.  Most clusters actually
lie in the bulge and disk regions in the regions we have excluded from analysis,
but a small sample of clusters can be constructed with small statistical errors.
As measured by agreement with parallax estimates from the catalog of \citet[][2010 edition]{harris96},
this correction scheme does not result in any clear improvement, though it does not
worsen the results either.
There are probably several reasons for this.
The waffle pattern most likely has only a minor effect on the mean parallaxes
of these clusters, because most of the signal comes from brighter stars for which 
the waffle pattern is weaker.
With regard to the large-scale field,
none of our sample clusters were in regions of extremely low or high bias, 
so the potential gain in accuracy was not great.
In addition, we already concluded the typical improvements offered by
the large-scale correction field are currently marginal at best.
There may be other systematic effects we are unaware of that apply to high-density
regions such as globular clusters.

We have written a Python package to estimate the parallax bias as a
function of the sky coordinates, constructed directly from the map in
Figure~\ref{fig.allskymaps_flattened} and its
relatives.\footnote{{\tt github.com/fardal/gaiadr2parbias}}
As shown by
the test versus globular clusters, these results are best used to
characterize rather than correct the systematic errors in the
parallax.  We note the bias estimates do not include the magnitude
dependence of the waffle pattern, since it is unclear how to combine
this with larger-scale variation of the parallax bias.  Even so, the
results from this package may be useful, for example, in testing whether a
particular object or sample is in a region of especially low or high bias.

\section{Connection to proper motion systematics}
\label{sec.pm}
So far we have focused on the parallax, because the information about
this quantity carried by the general stellar population is far greater
than for the proper motion.  The spread in proper motion at a given
color and magnitude is roughly ten times higher than the spread in
parallax.  Thus it is hard to observe the spatial patterns in
DR2's proper motion systematic errors, except in the regions with the highest
stellar density, which mostly means those at low latitude.
Still, it is worth seeing whether any of the
patterns we have found in the parallax help us understand the proper
motion systematics better.

\begin{figure}[t!]
\centerline{\includegraphics[width=180mm]{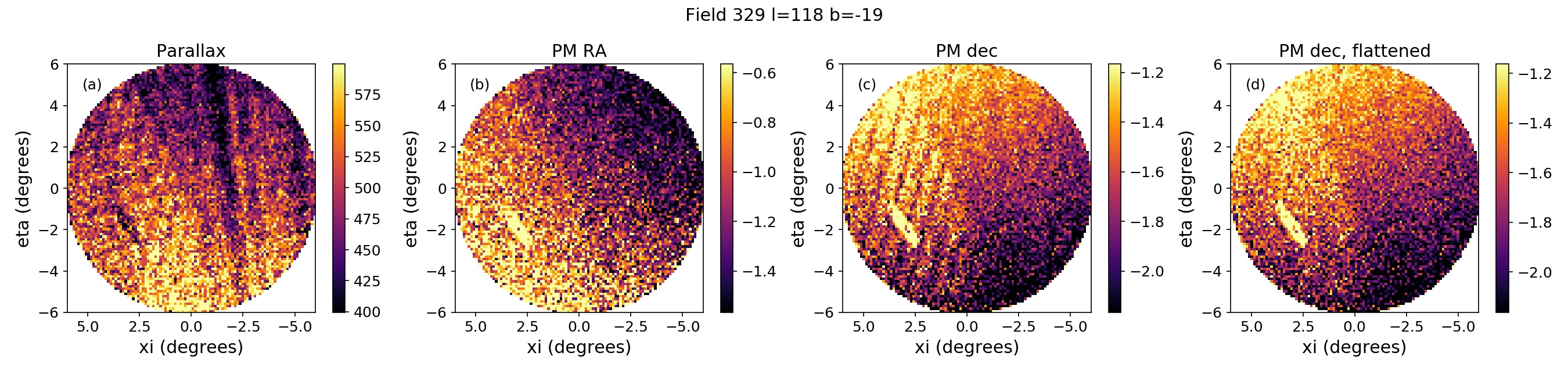}}
\caption{
  \label{fig.pm}
  Bias maps of a low-latitude field containing the M31 galaxy (the oval feature at lower left).
  Panel (a): weighted average parallax (colorbar labels in units of $\muas$;
  Panel (b): PM in right ascension (in units $\masyr$);
  Panel (c): PM in declination (in units $\masyr$);
  Panel (d): PM in declination, minus fit to waffle pattern.
  It can be seen that the subtraction flattens much of the systematic pattern,
  but does not fully handle sharp edge features.
}
\end{figure}

Figure~\ref{fig.pm} shows maps of parallax (panel (a)) and proper motion
(panels (b) and (c)) for 
the circular field in our sample that contains the M31 galaxy.
The motion of this galaxy is of great scientific interest, and
\citet{vdmarel19} already measured a proper motion
$(65 \pm 18, -57 \pm 15) \muasyr$ in right ascension (RA) and declination
using \gaia\ DR2 (neglecting systematic errors).
As with the parallax, we weight the 
proper motion using a measure of its spread as a function of color and magnitude.
The spread is again minimized for stars near the main sequence turnoff,
but is not as variable with magnitude.
Clearly, waffle patterns are present for the proper motion as well as
the parallax, as was previously demonstrated for the LMC.

Inspecting a number of low-latitude fields, it appears
the dominant Fourier modes of the proper motion waffle pattern
usually share alignments with those of the parallax.
We construct a smoothed version of the proper motion field in the
following way.  We again fit the parallax using our peak-aware
smoothing technique, and record the locations of the dominant Fourier
modes.  We then use these mode locations as the locations of peaks in
the proper motion, rather than searching for them anew, and then
smooth the proper motion.  Panel (d) of Figure~\ref{fig.pm}
shows the result of subtracting this smooth fit from the proper motion
map.  Clearly this technique largely flattens the stripe
features.  It does not fully flatten the sharp edge feature visible
in the last two panels, which makes sense since this smoothing
technique was not designed to fit sharp edges.
More testing would be required to assess the utility of this method,
and we present it here only to suggest the feasibility of flattening
systematic biases in later work.

In \citet{vdmarel19}, we accounted for systematic errors using a statistical
assessment of the errors based on \citet{lindegren18} and \citet{helmi18}.
In particular, we roughly replicated the waffle pattern observed in the LMC,
and estimated the effects it could have on the measured M31 proper motion.  
While the resulting estimate turns out to be reasonable, 
future work could take advantage of the specific pattern
exhibited by the general stellar population in Figure~\ref{fig.pm}.
The small-scale waffle pattern has rms $75 \muasyr$ variation in the neighborhood
of M31. This translates
to significant local velocity rms offsets of $>250 \kms$ at M31's distance,
although M31 is large enough that this partially averages out.
Accounting for the large-scale step feature in the PM systematic
that is shown in Figure~\ref{fig.pm}
could also be helpful to reduce systematic error in M31's proper motion.

We have noticed that in many fields,
the waffle pattern in RA proper motion 
emphasizes horizontal linear features (aligned along the RA direction
in this map projection), while the proper motion in declination emphasizes
vertical linear features (aligned along the declination direction).
This tendency is visible in the two central panels of Figure~\ref{fig.pm}.
In Fourier space, we find the largest proper motion amplitude vectors are
directed perpendicular to the wavenumber.  This implies a dominant apparent
motion along the streaks, or a divergence-free flow, rather than a
curl-free motion across them.  Qualitatively, we observe a similar 
behavior in some of the long, narrow scar features that are present in
the proper motion: the systematic offset compared to the scar's surroundings
appears directed along the streak, rather than across it.
We are only able to measure the proper motion systematic error at low Galactic
latitude, so we cannot say for sure whether these patterns persist throughout the sky.

\section{Conclusions}
\label{sec.conclusions}
In this paper we have probed the systematics of \gaia\ parallaxes using
the overall stellar population in \gaia\ DR2.  We have shown that the
the information in the stellar population,
dominated by turnoff stars with $G \sim 17$,
is sufficient to detect a variety of bias features of different
morphology and spatial scales.  Just as clearly,
it is difficult to completely separate out astrophysical effects
such as dust clouds and variation of the stellar population over the sky,
so our results must be viewed with a fair degree of caution.  We
have shown that the small-scale variations we term the ``waffle
pattern'', previously illustrated in isolated patches of the sky,
extend across the whole sky though with varying form and
amplitude.  We also demonstrated that the same pattern affects the
parallaxes of other types of sources, though not always to the same
degree, and demonstrated and quantified the decline of its impact with
increasing source brightness.  We have also detected higher-amplitude
elongated ``scar'' features, as well as large-scale regions with
parallax offsets.  While some of these features were shown previously,
the view obtained from our maps is both global and high resolution.
The various parallax bias features we have detected may have 
implications for other studies of the source parallaxes in \gaia,
particularly those using spatially compact groups of sources.
At this point, our work is more successful in 
describing the various systematic errors
than in offering a practical means to correct them.  
With a sample of globular clusters, we have {\em not} yet clearly
demonstrated a correction scheme
that markedly improves on the raw DR2 parallaxes.
This may owe to inaccuracy in the bias estimation, or to 
additional systematics that we have not yet identified.
Our short investigation of bias in the proper motion 
suggests a close relationship to the parallax bias, although
more work would be required to quantify the systematic error
in proper motion, or produce a useful correction scheme.  

The upcoming \gaia\ data releases should be exciting times when the
line separating knowledge from ignorance is suddenly pulled back.
We expect the data releases should markedly improve on both statistical
and systematic parallax errors, greatly expanding the volume accessible
to geometric distance determinations.
However, we do not yet know how the two sources of error will compare to each other, and it may take some time to assess this.
We intend to apply the tests in this paper, based on large numbers of Galactic stars,
to \gaia\ EDR3 and future data releases.
We expect these tests to be useful in surveying the new boundaries of our knowledge.

\acknowledgments

We thank Knut Olsen for help in starting this investigation and for many helpful comments.
We also acknowledge help from Stefano Casertano, Sihao Cheng, and Timo Prusti.
This work has made use of data from the European Space Agency (ESA) mission
{\it Gaia} (\url{www.cosmos.esa.int/gaia}), processed by the {\it Gaia}
Data Processing and Analysis Consortium (DPAC,
\url{www.cosmos.esa.int/web/gaia/dpac/consortium}),
and public auxiliary data provided by ESA/Gaia/DPAC
as obtained from the publicly accessible ESA Gaia SFTP.
Funding for the DPAC
has been provided by national institutions, in particular the institutions
participating in the {\it Gaia} Multilateral Agreement.
\software{numpy, scipy, scikit-learn, ipycache, astropy, astropy\_healpix, astroquery,
galstreams, IPAC-Montage, sfdmap}

\bibliography{autolib,gaia,new}{}
\bibliographystyle{aasjournal}
\end{document}